\documentclass[conference]{IEEEtran}
\pdfoutput=1
\IEEEoverridecommandlockouts

\usepackage{amsmath}
\usepackage{graphicx}
\usepackage{textcomp}
\usepackage{xcolor}
\usepackage[10pt]{moresize}
\usepackage{multirow}

\usepackage{amsthm}
\theoremstyle{definition}

\usepackage{algorithm}
\usepackage{algorithmicx}
\usepackage{algpseudocode}
\usepackage{comment}
\algdef{SE}[DOWHILE]{Do}{doWhile}{\algorithmicdo}[1]{\algorithmicwhile\ #1}%

\usepackage{epsfig}

\usepackage[font=footnotesize,skip=6pt]{caption}
\usepackage{subcaption}
\usepackage{url}

\def\BibTeX{{\rm B\kern-.05em{\sc i\kern-.025em b}\kern-.08em
    T\kern-.1667em\lower.7ex\hbox{E}\kern-.125emX}}
\begin{document}

\title{Timing Analysis Agent: Autonomous Multi-Corner Multi-Mode (MCMM) Timing Debugging with Timing Debug Relation Graph}

\author{\IEEEauthorblockN{Jatin Nainani}
\IEEEauthorblockA{\textit{NVIDIA} \\
Santa Clara, CA, USA \\
jnainani@nvidia.com}
\and
\IEEEauthorblockN{Chia-Tung Ho}
\IEEEauthorblockA{\textit{NVIDIA Research} \\
Santa Clara, CA, USA \\
chiatungh@nvidia.com}
\and
\IEEEauthorblockN{Anirudh Dhurka}
\IEEEauthorblockA{\textit{NVIDIA} \\
Santa Clara, CA, USA \\
adhurka@nvidia.com}
\and
\IEEEauthorblockN{Haoxing Ren}
\IEEEauthorblockA{\textit{NVIDIA Research}  \\
Austin, TX, USA \\
haoxingr@nvidia.com}
}

\maketitle

\begin{abstract}
\textcolor{black}{
Timing analysis is an essential and demanding verification method for Very Large Scale Integrated (VLSI) circuit design and optimization. In addition, it also serves as the cornerstone of the final sign-off, determining whether the chip is ready to be sent to the semiconductor foundry for fabrication. 
Recently, as the technology advance relentlessly, smaller metal pitches and the increasing number of devices have led to greater challenges, and longer turn-around-time for experienced human designers to debug timing issues from the Multi-Corner Multi-Mode (MCMM) timing reports. 
As a result, an efficient and intelligent methodology is highly necessary and essential for debugging timing issues and reduce the turnaround times.}

\textcolor{black}{Recently, Large Language Models (LLMs) have shown great promise across various tasks in language understanding and interactive decision-making, incorporating reasoning and actions. 
In this work, we propose a timing analysis agent, that is empowered by multi-LLMs task solving, and incorporates a novel hierarchical planning and solving flow to automate the analysis of timing reports from commercial tool. 
In addition, we build a Timing Debug Relation Graph (TDRG) that connects the reports with the relationships of debug traces from experienced timing engineers.
The timing analysis agent employs the novel Agentic Retrieval Augmented Generation (RAG) approach, that includes agent and coding to retrieve data accurately, on the developed TDRG.
In our studies, the proposed timing analysis agent achieves an average 98\% pass-rate on a single-report benchmark and a 90\% pass-rate for multi-report benchmark from industrial designs, demonstrating its effectiveness and adaptability.
}

\end{abstract}


\section{Introduction}
Very Large Scale Integration (VLSI) circuit design and optimization require extensive timing analysis and debugging to verify functional correctness. Today, timing analysis serves as the cornerstone of the final sign-off process, determining whether the design is ready for chip fabrication. As technology has advanced beyond 5 {\em nm}, the growing number of transistors, increasingly complex circuit designs, and the amplified effects of cross-talk due to smaller metal pitches have introduced significant challenges. These factors lead to longer debugging times for experienced timing analysis engineers who must address timing issues based on Multi-Corner Multi-Mode (MCMM) timing reports.

Recent advances have explored the use of Large Language Models (LLMs) as agents for interactive decision-making~ \cite{yao2022react}. 
\textcolor{black}{However, MCMM timing report analysis remains an unsolved and challenging task because the context of multi-report analysis exceeds the token limits of current state-of-the-art LLMs and requires a wide variety of queries across different types of timing reports. }
Although recent works~\cite{bairi2024codeplan, liu2024stall+, luo2024repoagent} have proposed repository-level agent frameworks for tasks like code planning and coding assistance, these solutions lack the domain knowledge necessary to handle various report types (e.g., path delay, cross-talk, logic constraints) simultaneously.
Consequently, an efficient and intelligent methodology is essential to streamline the debugging of timing issues and reduce turnaround time.

\begin{figure}[t]
    \centering
    \includegraphics[width=1.005\columnwidth]{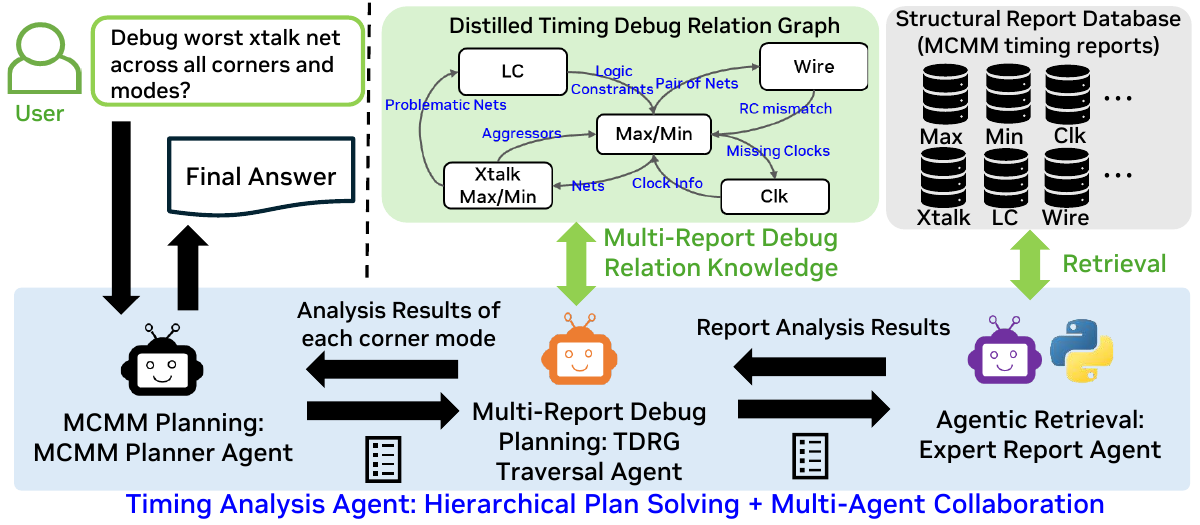} 
    \caption{An illustration of Timing Analysis Agent, which integrates hierarchical planning, multi-agent collaborations, and the novel distilled Timing Debug Relation Graph to solve MCMM timing task.}
    \label{AgentIllustrationFig}
\end{figure}

\textcolor{black}{In this paper, we propose a novel and intelligent Timing Analysis Agent, which integrates multi-LLM task-solving, hierarchical planning, and the expertise of experienced timing engineers to generate insightful analysis and debug timing issues with high accuracy and minimal variance, as shown in Fig.~\ref{AgentIllustrationFig}. 
The proposed methodology aims at solving the general timing analysis and debug problems of MCMM timing reports.}
The contributions of our work include:
\begin{enumerate}
    \item We propose a novel Timing Analysis Agent that integrates hierarchical plan solving and multi-agent collaboration to automate the analysis of MCMM timing reports generated by timing verification tools (e.g., Static Timing Analysis (STA), variation analysis, etc.). 
    \item We distill the debugging trace into a Timing Debug Relation Graph (TDRG), which connects individual reports with debug knowledge from experienced timing engineers. The proposed Timing Analysis Agent leverages the TDRG to dynamically traverse and retrieve information on nets, paths, or instances across multiple timing reports to solve tasks.
    \item We introduce a novel variation of Agentic Retrieval Augmented Generation (RAG), which leverages the coding capabilities of LLMs to retrieve necessary timing information from the reports, excluding irrelevant data. This approach ensures the retrieval is adaptable to a wide range of queries.
    \item We develop the MCMM planner agent, the TDRG traversal agent, and the expert report agent to solve tasks hierarchically—starting with mode planning, followed by multi-report traversal, single-report information retrieval, and culminating in providing the final answer.
    \item We validate the effectiveness of our proposed Timing Analysis Agent through extensive experiments. The results show that our approach outperforms other RAG techniques by more than 46\% and achieves a 98\% pass-rate on average for single-report benchmark and a 90\% pass-rate for multi-report benchmark.
\end{enumerate}

The remaining sections are organized as follows: Section~\ref{LiteratureReview} reviews the related works on leveraging LLMs for multi-file analysis, retrieval, and applications in Electronic Design Automation (EDA) field. Section~\ref{BackGround} discusses the timing analysis tasks in detail. Then, we introduces the proposed Timing Analysis Agent methodology in Section~\ref{TimingAgent}. Section~\ref{Experiments} presents our main experiment and sensitivity studies. Lastly, Section~\ref{Conclusion} concludes the paper.

\vspace{-0.1cm}
\section{Literature Review} \label{LiteratureReview}
We review related works on multi-file analysis, structural data analysis, and LLM applications in EDA below.

\vspace{-0.1cm}
\subsection{Multi-file Analysis and Editing}

Bairi et al. \cite{bairi2024codeplan} implemented repository-level coding through a task-agnostic framework called CodePlan, which frames coding as a planning problem. 
This system synthesizes a multi-step plan of interdependent edits across a repository by combining static analysis, change-impact assessment, and adaptive planning.
Their method allows LLMs to handle complex tasks such as package migration and temporal code edits, outperforming simpler oracle-guided systems by ensuring consistency throughout the repository. 
Liu et al. \cite{liu2024stall+} extended repository-level code completion with STALL+, which integrates static analysis techniques to enhance LLM performance. By leveraging static analysis for dependency identification, STALL+ improves context comprehension and boosts LLM accuracy on large repositories. Luo et al. \cite{luo2024repoagent}, with RepoAgent, focused on repository-level documentation generation. Their framework uses LLMs to analyze code and generate comprehensive documentation across entire repositories, further demonstrating that LLMs can manage complex multi-file tasks.
These works illustrate the growing capability of LLM agents in handling repository-wide tasks. However, the domain-specific complexity of MCMM timing reports, which requires expert trace distillation and cross-corner reasoning, remains a challenge. 

\vspace{-0.1cm}
\subsection{Structured Data Analysis with Large Language Models}

The application of LLMs to structured data remains a significant challenge due to the complexity and scale of such datasets.
Recently, several works have focused on efficient retrieval methodologies, Ziletti et al.~\cite{ziletti2024retrieval} showcase how LLM, using the retrieval augmented text-to-SQL technique, can be applied to answer epidemiological questions using EHR and claims data. 
Dong and Wang~\cite{dong2024large} review the current state of LLMs for tabular data, highlighting progress in fine-tuning and data representation, while also emphasizing limitations in handling large, heterogeneous datasets without significant adaptation. 
Jiang et al.~\cite{jiang2023structgpt} propose StructGPT to improve reasoning over structured data through prompt engineering. 
These methods usually assume a relational database is already present and therefore can not be directly applied to efficiently retrieve data from MCMM timing reports.
Additionally, Sui et al.~\cite{sui2024table} benchmark LLMs on tabular data, revealing that while LLMs are promising, they underperform on complex datasets typical in EDA. These findings suggest that despite recent advancements, LLMs require further refinement to be effective in large-scale, multi-file analysis tasks.

\begin{figure}[t]
    \centering
    \includegraphics[width=0.8\linewidth]{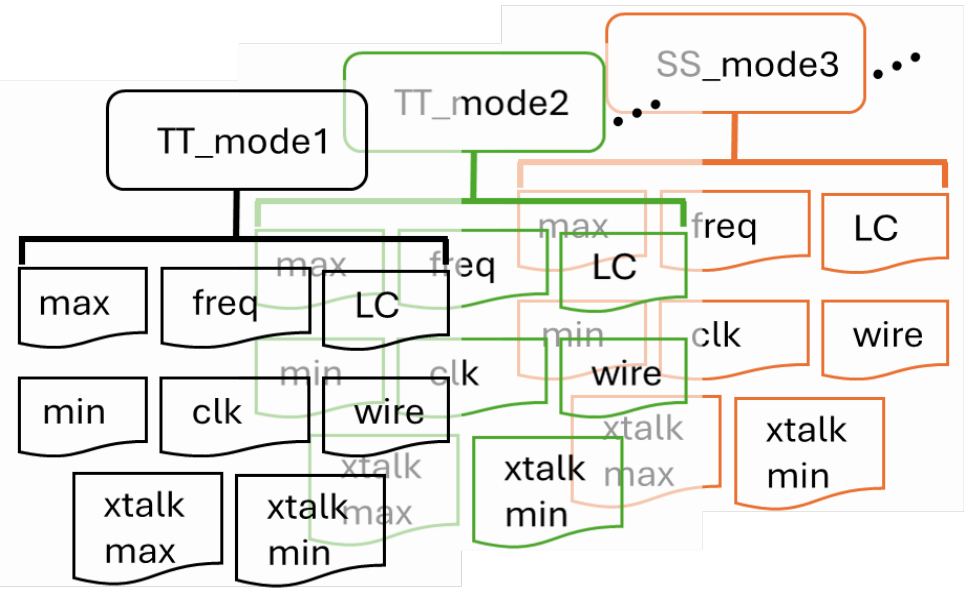} 
    \vspace{-0.25cm}
    \caption{Hierarchical relations of the MCMM timing reports. There are multiple PVT corners and modes, like TT\_mode1, SS\_mode3, etc. Each of corner and mode has max, min, xtalk\_max, xtalk\_min, clk, freq, LC, and wire reports.}
    \label{TimingReportHierarchyFig}
\end{figure}

\begin{figure*}[t]
    \centering
    \includegraphics[width=0.85\linewidth]{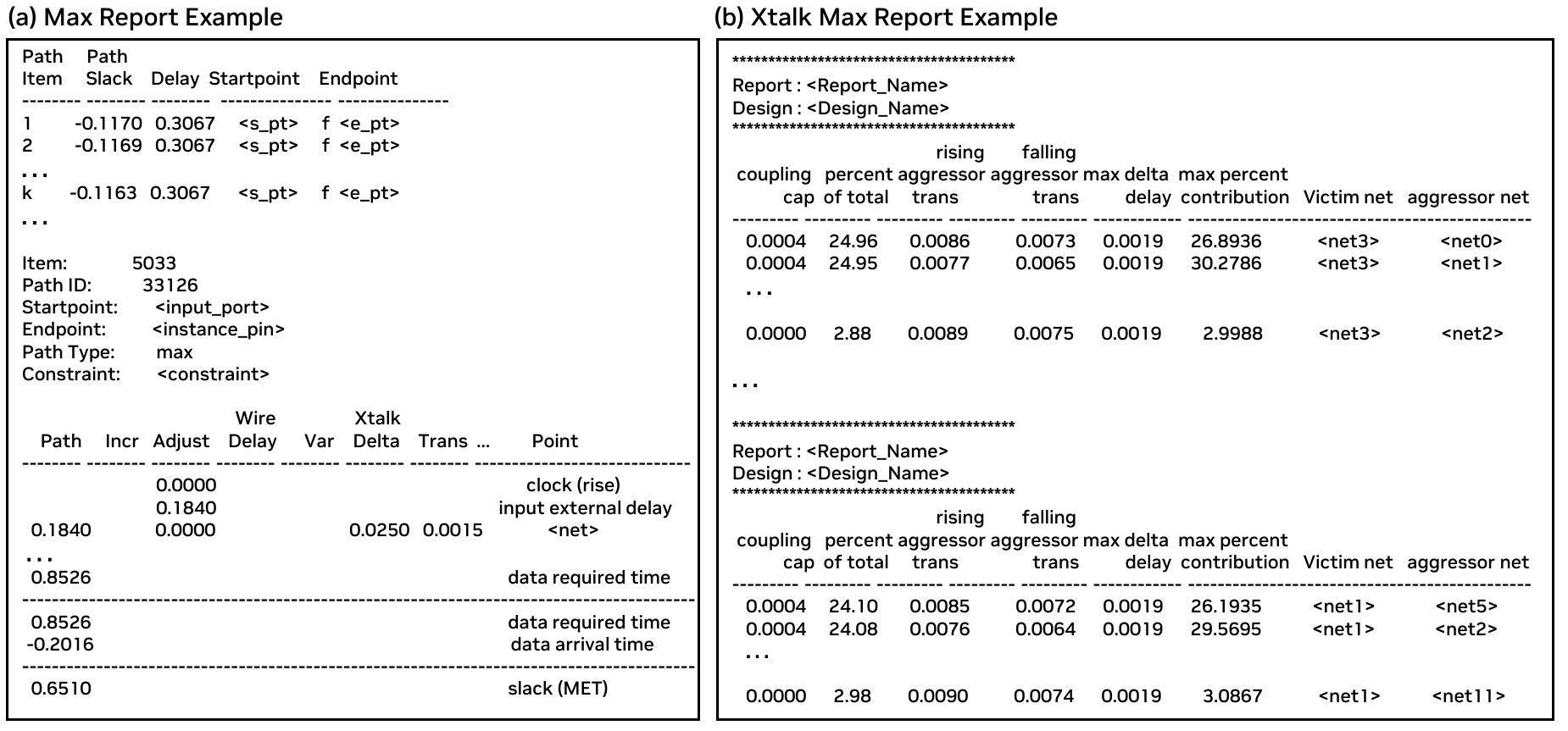} 
    \vspace{-0.25cm}
    \caption{Examples of max and xtalk\_max timing report for a specific corner and mode.}
    \label{TimingReportExampleFig}
\end{figure*}

\vspace{-0.1cm}
\subsection{Large Language Models in Electronic Design Automation}

The integration of Large Language Models (LLMs) in EDA has shown promise across various tasks. Chang et al.~\cite{chang2024data} developed a framework that utilizes data augmentation to fine-tune LLMs for chip design, particularly in generating Verilog code and EDA scripts. 
This framework improves the alignment between natural language and hardware description languages, enhancing the model's ability to generate and repair Verilog code. Ho and Ren~\cite{ho2024large, ho2024verilogcoder} explored the application of LLMs in optimizing standard cell layout design and Verilog coding, demonstrating the potential for integrating domain-specific knowledge into layout tasks. Similarly, Liu et al.~\cite{liu2024layoutcopilot} introduced LayoutCopilot, a multi-agent collaborative framework for interactive analog layout design, which leverages LLMs to streamline the design process.
Additionally, Wu et al.~\cite{wu2024eda} presented ChatEDA, an approach that uses LLMs to autonomously generate and interact with EDA scripts within the OpenROAD environment, showcasing the potential for automating EDA workflows. 
While these advancements highlight the capabilities of LLMs in EDA, the focus has largely been on isolated tasks such as layout optimization, script generation, and Verilog code synthesis.

In summary, the prior related works can not be applied for complex timing debugging and analyses of MCMM timing reports since it requires cross-referencing and multi-hop reasoning across reports with expert domain knowledge.

\vspace{-0.2cm}
\section{Background} \label{BackGround}

Here, we introduce the background of timing reports and the selected real world timing tasks from expert timing engineers. We create the benchmark based on the selected timing tasks for evaluation in the experiments.

\vspace{-0.2cm}
\subsection{Timing Reports}
Timing analysis tools cover gate-level STA, detailed transistor-level STA, and variation analysis for VLSI designs including intricate custom designs. 
Silicon failures in advanced technologies like FinFET can be costly, making rigorous signoff analysis crucial to avoid critical timing and noise issues. 
Timing analysis is typically run across the entire circuit, which contains multiple corners.
Each corner includes modes like read, write, scan, etc. 
For each corner and mode, there are max, min, xtalk\_max, xtalk\_min, clk, freq, Logic Constraints (LC) and wire types of report for timing engineer to debug the timing issues.

Fig~\ref{TimingReportHierarchyFig} illustrates the hierarchical relations of the MCMM timing reports. 
Fig~\ref{TimingReportExampleFig} provides examples of max, and xtalk\_max timing reports that usually contain more than 16,000 timing paths for an industrial design.
In Fig~\ref{TimingReportExampleFig}, the format and attributes of the timing reports vary, posing challenges for LLMs when retrieving the necessary information for different timing tasks.

\vspace{-0.2cm}
\subsection{Timing Tasks} 
We outline the timing tasks selected by experienced timing engineers to create a benchmark from industrial designs for evaluating performance, categorizing them into single-report and multi-report tasks.

\noindent {\bf Single-Report Tasks}: 
Single-report tasks require domain knowledge of the type of report being analyzed and an understanding of its intricate structure. We divide single-report tasks into two categories and provide examples below.

\noindent {\em Query or group timing paths based on specific criteria}: These tasks commonly involve retrieving paths between a "startpoint" and an "endpoint," identifying paths with specific "Constraint" values, or finding paths that start with specific edges across the data and clock arcs. Some tasks require performing complex queries to retrieve specific sets of problematic paths that need further investigation, while others focus on deeply examining individual paths to identify issues.

\noindent{\em Perform mathematical or string operations on a group of paths}: These tasks usually involve performing operations (e.g., max, min, avg) on columns or high-level attributes. For example, one might compare the worst values of different attributes across multiple paths or find the top-k paths with timing issues.

\begin{figure*}[t]
    \centering
    \includegraphics[width=\linewidth]{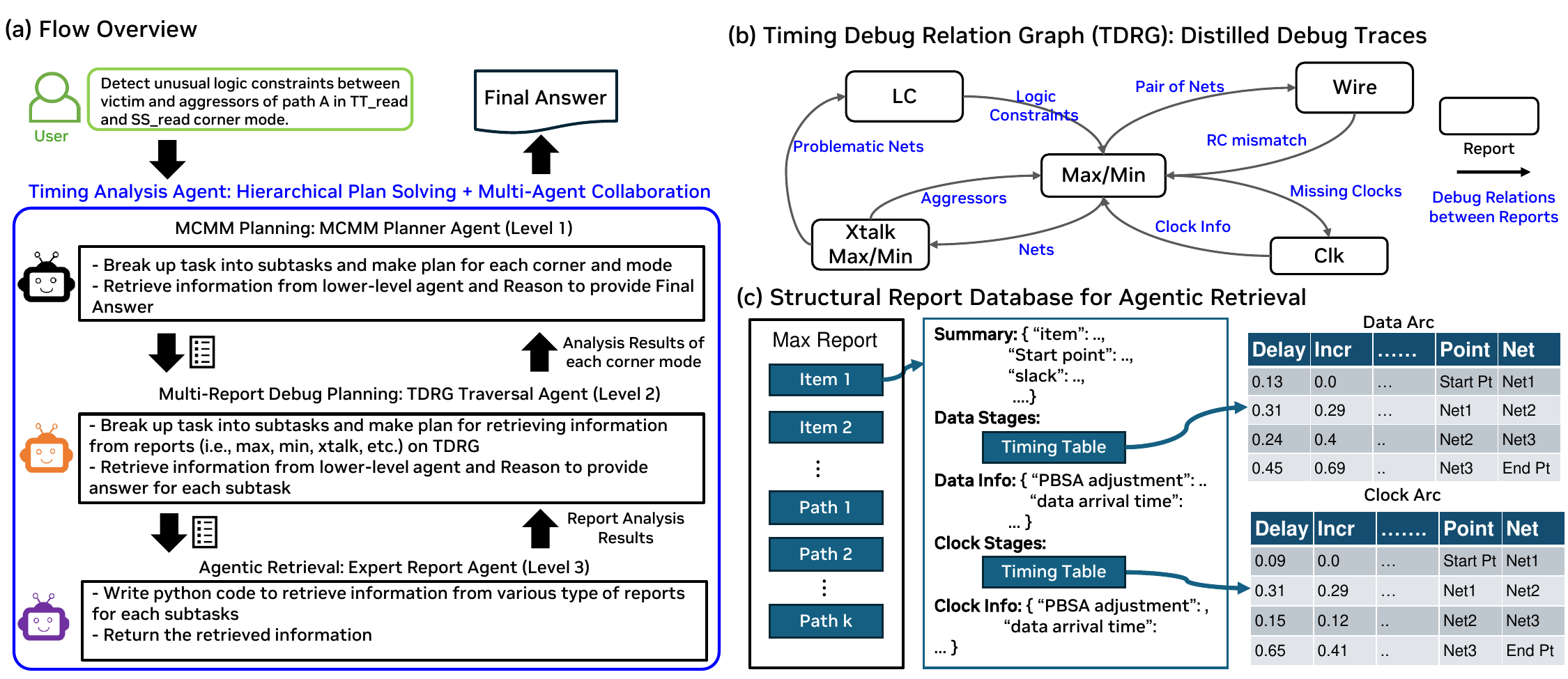} 
    \caption{(a) Flow overview of Timing Analysis Agent with hierarchical plan solving, and multi-agent collaboration. (b) Timing Debug Relation Graph (TDRG) from distilled debug traces from experienced timing engineer. (c) Structural report database for coding agentic retrieval.}
    \label{OverallFlowFig}
\end{figure*}

\noindent{\bf Multi-Report Tasks}:
The process of completing multi-report tasks often involves gathering data from various types of reports (e.g., max, xtalk\_max, etc.) and performing reasoning to derive the final conclusion. 
For example, detecting missing clock signals in a timing path for a specified corner and mode requires cross-reference the clock signals in the clk report and timing tables in the max report.
Another example involves identifying unusual RC values between victim and aggressor nets in a timing path.
This first requires finding the most significant "aggressor" and "victim" nets in the xtalk report. 
Then, the next step is investigating the high RC mismatch and any abnormal logic constraints between the identified aggressor and victim nets in the wire and LC reports, respectively. 
Here, the RC mismatch is determined by subtracting the worst-case RC values. 
To perform analysis across MCMM, the multi-report task must be conducted in various corners and modes. The results are then summarized, providing explanations for any potential causes of identified problems across all corners and modes.




\vspace{-0.2cm}
\section{Timing Analysis Agent} \label{TimingAgent}

The Timing Analysis Agent integrates hierarchical plan solving, and multi-agent collaboration, as shown in Fig~\ref{OverallFlowFig}(a). 
Given a user task (e.g., violation search or identifying nets with significant cross-talk impact), the MCMM Planner Agent first decomposes the task into a list of corner and mode specified sub-tasks. Then, the TDRG Traversal Agent makes a report retrieval plans, calls Expert Report Agents to retrieve information, and summarizes the fetched net delay, path delay, and cross-talk data from multiple reports on the proposed novel distilled TDRG as shown in Fig~\ref{OverallFlowFig}(b). Finally, the MCMM planner agent compiles the responses of sub-tasks, and returns the Final Answer. 
We introduce the structural report database and the framework in a bottom-up approach. 

\vspace{-0.2cm}
\subsection{Structural Report Database}
Given the MCMM timing reports as input, we utilize the original structure of each timing report to construct a structured report database, organized using nested dictionaries for each report type. Figure~\ref{OverallFlowFig}(c) illustrates an example of a structured database for a max report.
In this database, the max report is organized into key attributes: "Summary," "Data Info," "Clock Info," "Data Stages," and "Clock Stages," allowing for efficient and informed access. 
"Summary" attribute contains high-level path information, such as startpoint, endpoint, path ID, slack, etc.
The "Data Info" and "Clock Info" attributes summarize timing arc information, including PBSA adjustments and data arrival times.
Finally, the detailed timing table for each timing arc is stored in the "Data Stages" and "Clock Stages" as shown in the rightmost part of Figure~\ref{OverallFlowFig}(c). 

\vspace{-0.2cm}
\subsection{Timing Debug Relation Graph (TDRG)}
\textcolor{black}{We construct the TDRG based on a distilled debugging trace provided by experienced timing engineers, enabling the agent to develop a plan for retrieving information from multiple timing reports.
In the TDRG, nodes represent the reports, while edges define the relationships between the debug information, as shown in Figure~\ref{OverallFlowFig}(b).
For each node, we describe the timing attributes, type of timing table, and the usage of the corresponding report. For example, the description of a wire report node includes details such as the "Worst\_RC" attributes and the calculation of RC mismatch. 
On the other hand, the edge descriptions capture distilled knowledge from the debugging trace, describing the relationships between different types of reports. 
For instance, the edge relationship between xtalk and LC reports is "find the logic constraints on the aggressor and victim nets."}


\begin{figure}[t]
    \centering
    \includegraphics[width=\linewidth]{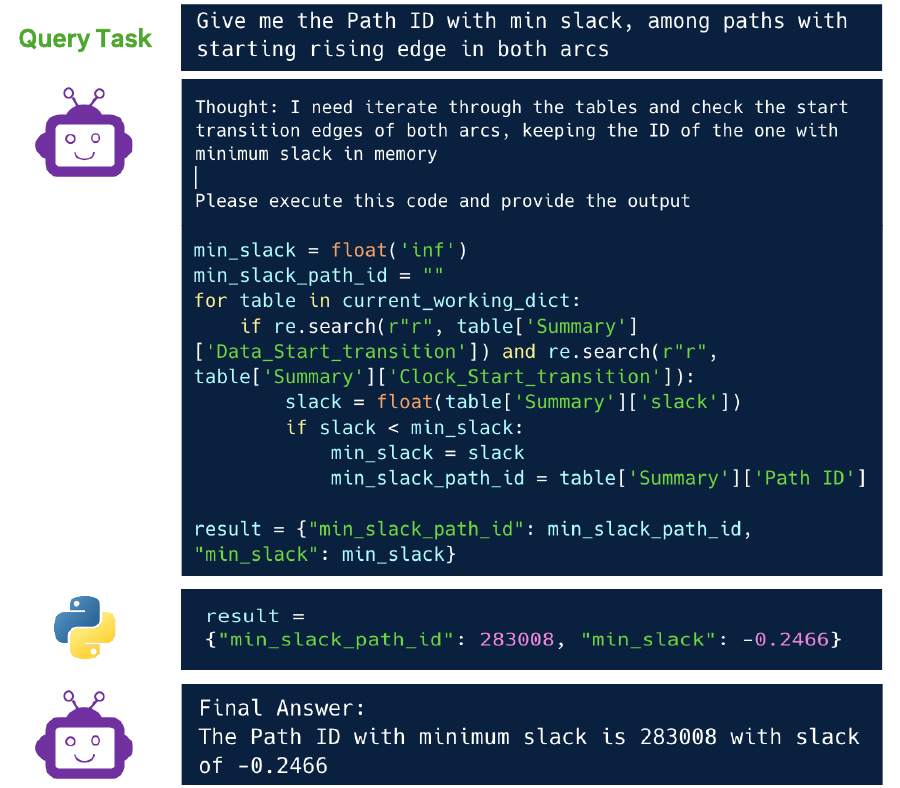} 
    \caption{An example of Expert Report Agent writing Python code to retrieve path ID of the minimum slack in max report.}
    \label{ExpertReportAgentRetrievalExample}
\end{figure}

\vspace{-0.15cm}
\subsection{Expert Report Agent: Coding Agentic Retrieval (level 3)}

We develop a novel Expert Report Agent that leverages the coding capabilities of LLMs to retrieve the needed timing information without including extraneous data that could fill the context limit.
Inspired by the findings of Wang et al.~\cite{wang2024executable}, which demonstrate that agents capable of executing code outperform other methods, we designed the Expert Report Agent to tailor its functionality to each report type (e.g., max, crosstalk, etc.). 
The agent generates flexible and adaptable Python code to query the structural report database according to the specific task requirements. 

Figure~\ref{ExpertReportAgentRetrievalExample} illustrates an example where the Expert Report Agent writes Python code to retrieve the path ID of the minimum slack from the max report. Given the task query, the Expert Report Agent first writes the python code for execution. After retrieving the information, the Expert Report Agent summarizes the retrieved information and provides the answer to the query.



\vspace{-0.2cm}
\subsection{TDRG Traversal Agent: Multi-Report Debugging (level 2)}

The leverages the distilled TDRG to create a plan to retrieve information (i.e., unusual nets, aggressors, and logic constraints, etc.), and then summarizes the results to complete the task. 
The TDRG Traversal Agent calls Expert Report Agents to retrieve the unusual nets in max report, aggressors in xtalk max report, and logic constraints in LC report, respectively.

\begin{figure}[t]
    \centering
    \includegraphics[width=1.005\columnwidth]{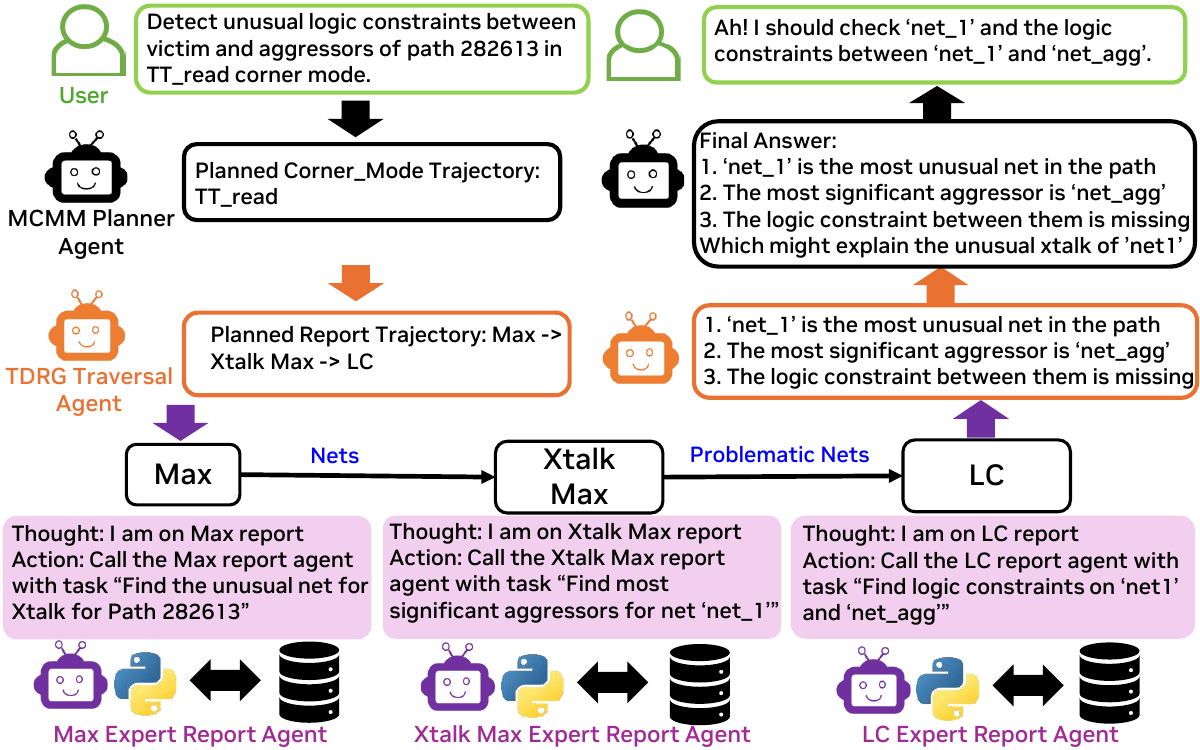} 
    \caption{Illustration of the hierarchical plan solving and multi-agent collaboration in solving the multi-report task.}
    \label{MultiLevelSolvingExampleFig}
\end{figure}

\vspace{-0.2cm}
\subsection{MCMM Planner Agent: MCMM Planning (level 1)}
The MCMM Planner Agent decomposes the user-provided MCMM timing task into sub-tasks for each corner and mode \textcolor{black}{based on the prompt from user}. Then, MCMM Planner Agent calls TDRG Traversal Agent to resolve each sub-task sequentially.
Finally, the MCMM Planner Agent collects the analysis results for each planned corner and mode from the TDRG Traversal Agent and provides the final answer to the user.

Fig~\ref{MultiLevelSolvingExampleFig} illustrates the hierarchical plan solving and multi-agent collaboration in solving the task "Detect unusual logic constraints between victim and aggressors of path 282613 in TT\_read corner mode" task. 
The MCMM Planner Agent makes a planned trajectory of mode and corner and instructs TDRG Traversal Agent for completing the task.
The TDRG Traversal Agent generates sub-tasks for multi-report retrieval.
The Expert Report Agents retrieve the unusual net, aggressor, and logic constraint, then return this information to TDRG Traversal Agent.
Lastly, the MCMM Planner Agent summarizes the analysis messages from TDRG Traversal Agent and provides the Final Answer to user.

\vspace{-0.1cm}
\section{Experimental Results} \label{Experiments}
Our work is implemented in Python and is built on top of the Autogen~\cite{wu2023autogen} multi-LLM agent framework.
We use Llama3~\cite{Llama3} for each agent. The temperature and top\_p parameters of the LLM are set to 0.3 and 1.0, respectively. 
We create a single-report benchmark to evaluate retrieval effectiveness and correctness, and a multi-report benchmark to assess the hierarchical planning and reasoning capabilities of the agent from industrial designs, as described in Section~\ref{BackGround}. The pass rates of the single-report and multi-report benchmarks are compared to the golden answer and evaluated by experienced human engineers. There are 10 tasks for each general task category in the single-report benchmark. Firstly, we demonstrate the effectiveness of the proposed coding agent retrieval on the single-report benchmark. Next, we present the pass rate of the proposed Timing Analysis Agent in multi-report debugging and reasoning. Lastly, we conduct a sensitivity study on the impact of information from TDRG on multi-report tasks.

\vspace{-0.1cm}
\subsection{Single-Report Experiment}

\begin{table}[]
\caption{Pass-rate (\%) of keyword search~\cite{jones2000probabilistic}, embed search~\cite{lewis2020retrieval}, hybrid search~\cite{HybridSearch}, functional agentic retrieval (Agentic Function), and proposed approach (Agentic Coding) on single-report benchmark table.}
\label{SingleBenchmarkTbl}
\vspace{-0.2cm}
\tabcolsep = 2pt
\centering
\scriptsize{
\begin{tabular}{|c|r|r|r|r|r|r|}
\hline
General Task Category                                                                                    & \multicolumn{1}{c|}{\#Tasks} & \multicolumn{1}{c|}{\begin{tabular}[c]{@{}c@{}}Keyword\\ \cite{jones2000probabilistic}\end{tabular}} & \multicolumn{1}{c|}{\begin{tabular}[c]{@{}c@{}}Embed\\ \cite{lewis2020retrieval}\end{tabular}} & \multicolumn{1}{c|}{\begin{tabular}[c]{@{}c@{}}Hybrid\\ \cite{HybridSearch}\end{tabular}} & \multicolumn{1}{c|}{\begin{tabular}[c]{@{}c@{}}Agentic \\ Function\end{tabular}} & \multicolumn{1}{c|}{\begin{tabular}[c]{@{}c@{}}Agentic \\ Coding \\ (proposed)\end{tabular}} \\ \hline
Check path for violation                                                                                 & 10                           & 0.0                                                                           & 0.0                                                                         & 0.0                                                                          & 100.0                                                                            & 100.0                                                                                        \\ \hline
\begin{tabular}[c]{@{}c@{}}Find worst case \\ \textless{}attribute\textgreater across paths\end{tabular} & 10                           & 0.0                                                                           & 0.0                                                                         & 0.0                                                                          & 90.0                                                                             & 100.0                                                                                        \\ \hline
\begin{tabular}[c]{@{}c@{}}Find worst case \\ \textless{}column\textgreater across paths\end{tabular}    & 10                           & 0.0                                                                           & 0.0                                                                         & 0.0                                                                          & 70.0                                                                             & 90.0                                                                                         \\ \hline
\begin{tabular}[c]{@{}c@{}}Check if a given path is \\ external or internal\end{tabular}                 & 10                           & 0.0                                                                           & 0.0                                                                         & 0.0                                                                          & 80.0                                                                             & 100.0                                                                                        \\ \hline
\begin{tabular}[c]{@{}c@{}}Which is the slowest \\ stage in the whole path\end{tabular}                  & 10                           & 0.0                                                                           & 0.0                                                                         & 0.0                                                                          & 20.0                                                                             & 100.0                                                                                        \\ \hline
\begin{tabular}[c]{@{}c@{}}Net with max crosstalk \\ delta in path\end{tabular}                          & 10                           & 0.0                                                                           & 0.0                                                                         & 0.0                                                                          & 30.0                                                                             & 90.0                                                                                         \\ \hline
\begin{tabular}[c]{@{}c@{}}Slew on the \textless{}net\textgreater \\ in path\end{tabular}                & 10                           & 0.0                                                                           & 0.0                                                                         & 0.0                                                                          & 0.0                                                                              & 100.0                                                                                        \\ \hline
\begin{tabular}[c]{@{}c@{}}Path goes through \\ \textless{}net\textgreater{}?\end{tabular}               & 10                           & 0.0                                                                           & 0.0                                                                         & 0.0                                                                          & 50.0                                                                             & 100.0                                                                                        \\ \hline
\begin{tabular}[c]{@{}c@{}}Data arc goes through \\ \textless{}clk\textgreater rising?\end{tabular}      & 10                           & 0.0                                                                           & 0.0                                                                         & 0.0                                                                          & 30.0                                                                             & 100.0                                                                                        \\ \hline \hline
Average                                                                                                  & 10                           & 0.0                                                                           & 0.0                                                                         & 0.0                                                                          & 52.0                                                                             & 97.8                                                                                         \\ \hline
\end{tabular}
}
\end{table}

We demonstrate the pass-rate of the novel coding agentic retrieval method and compare it with prior works using a single-report benchmark.
Each task is performed once, and the average pass rate for each general task category is calculated.
we compare the proposed coding agentic retrieval approach with 
Keyword Search~\cite{jones2000probabilistic}, Embedding search~\cite{lewis2020retrieval}, and a hybrid search~\cite{HybridSearch}.
In addition, we compare the proposed approach with functional agentic retrieval approach to further show the flexibility and effectiveness of the proposed coding agentic retrieval method for dynamic task content.
Basic functions (i.e., "check\_path\_for\_violation," "get\_specific\_attribute," and "get\_total\_column\_value") are implemented to extract information from the structural database of reports for the functional agentic retrieval approach. 

Table~\ref{SingleBenchmarkTbl} shows the pass-rate of the proposed coding agentic retrieval approach (i.e., Agentic Coding) and other baseline methods. 
Keyword Search~\cite{jones2000probabilistic}, Embedding search~\cite{lewis2020retrieval}, and hybrid search fail to retrieve the specific columns or values required for the task, as these unstructured approaches either retrieve entire timing tables or unrelated information. 
Compared to functional agentic approach (i.e., Agentic Function), the proposed coding agentic approach achieves 45.8\% higher pass-rate on average of all the general task categories since functional agentic approach is limited by predefined functions and cannot scale to the diverse types of queries that involve many combinations of basic column or value extractions.
On the other hand, the proposed coding agentic approach scales to various combinations of basic queries without requiring predefined functions for retrieval.
Overall, the proposed coding agentic approach achieves 98\% pass-rate on average for single-report benchmark.

\vspace{-0.2cm}
\subsection{Multi-Report Experiment}



\begin{table}[]
\caption{Pass-rate (\%) of the proposed approach, and a naive LLM planner without TDRG on multi-report benchmark table.}
\label{MultiReportTbl}
\vspace{-0.25cm}
\tabcolsep = 2pt
\centering
\scriptsize{
\begin{tabular}{|cc|c|r|r|}
\hline
\multicolumn{1}{|c|}{Task ID} & Task Description                                                                                                               & \begin{tabular}[c]{@{}c@{}}Required types \\ of reports\end{tabular} & \multicolumn{1}{c|}{\begin{tabular}[c]{@{}c@{}}LLM Planner \\ wo Graph\end{tabular}} & \multicolumn{1}{c|}{Proposed} \\ \hline
\multicolumn{1}{|c|}{M1}      & \begin{tabular}[c]{@{}c@{}}Find missing clk signals that \\ have no rise/fall information\end{tabular}                         & max, clk                                                             & X                                                                                    & V                             \\ \hline
\multicolumn{1}{|c|}{M2}      & \begin{tabular}[c]{@{}c@{}}Identify pairs of nets with \\ high RC mismatch\end{tabular}                                        & max, wire                                                            & X                                                                                    & V                             \\ \hline
\multicolumn{1}{|c|}{M3}      & \begin{tabular}[c]{@{}c@{}}Detect unusual constraints \\ between victim and its aggressors\end{tabular}                        & \begin{tabular}[c]{@{}c@{}}max, xtalk, \\ LC\end{tabular}            & X                                                                                    & V                             \\ \hline
\multicolumn{1}{|c|}{M4}      & \begin{tabular}[c]{@{}c@{}}Identify unusual RC values \\ between victim and its aggressors\end{tabular}                        & \begin{tabular}[c]{@{}c@{}}max, wire, \\ xtalk, LC\end{tabular}      & X                                                                                    & V                             \\ \hline
\multicolumn{1}{|c|}{M5}      & \begin{tabular}[c]{@{}c@{}}Find the constraints of slowest \\ stages with highest RC values\end{tabular}                       & \begin{tabular}[c]{@{}c@{}}max, wire, \\ xtalk, LC\end{tabular}      & X                                                                                    & V                             \\ \hline
\multicolumn{1}{|c|}{M6}      & \begin{tabular}[c]{@{}c@{}}Compare each timing table for \\ number of stages, point values \\ and timing mismatch\end{tabular} & max                                                                  & X                                                                                    & X                             \\ \hline
\multicolumn{1}{|c|}{M7}      & \begin{tabular}[c]{@{}c@{}}Task M2 and Task M3 for \\ specific stages in list of paths\end{tabular}                            & \begin{tabular}[c]{@{}c@{}}max, wire, \\ xtalk, LC\end{tabular}      & X                                                                                    & V                             \\ \hline
\multicolumn{1}{|c|}{M8}      & Task M1 across all modes                                                                                                       & max, clk                                                             & X                                                                                    & V                             \\ \hline
\multicolumn{1}{|c|}{M9}      & Task M3 across all modes                                                                                                       & \begin{tabular}[c]{@{}c@{}}max, xtalk, \\ LC\end{tabular}            & X                                                                                    & V                             \\ \hline
\multicolumn{1}{|c|}{M10}     & \begin{tabular}[c]{@{}c@{}}Task M2 and Task M3 across \\ all modes\end{tabular}                                                & \begin{tabular}[c]{@{}c@{}}max, wire, \\ xtalk, LC\end{tabular}      & X                                                                                    & V                             \\ \hline \hline
\multicolumn{2}{|c|}{Average pass-rate (\%)}                                                                                                                   & -                                                                    & 0.0                                                                                  & 90.0                          \\ \hline
\end{tabular}
}
\end{table}

We study the problem solving, and reasoning abilities of the proposed Timing Analysis Agent on multi-report benchmark in Table~\ref{MultiReportTbl}, which includes task descriptions and the required types of reports for each task. 
As we are first to approach the multiple timing reports debugging and analysis tasks, we compare the proposed Timing Analysis Agent with a naive LLM planner that does not utilize TDRG.
The proposed Timing Analysis Agent achieves a 90\% pass-rate, whereas the naive LLM planner struggles to solve any of the multi-report tasks due to its lack of debug trace integration between timing reports, which is necessary for creating accurate plans.
In summary, the proposed Timing Analysis Agent successfully and effectively resolve the multi-report tasks using the novel hierarchical plan solving process and distilled TDRG from debug traces of experienced timing engineers.


\subsection{Sensitivity Study of TDRG}
We conducted an extensive study to evaluate the impact and sensitivity of node descriptions and edge information on TDRG across five multi-report tasks, each specified by a particular corner and mode (i.e., from Task M1 to Task M5) as detailed in Table~\ref{MultiReportTbl}. 
We vary the node description and edge information for agent at level 2, as shown in Figure~\ref{OverallFlowFig}(a), to make a plan for solving multi-report tasks. 
The different settings are as follows:
\begin{enumerate}
    \item {\em Set1}: No information about reports and their relations.
    \item {\em Set2}: Limited node description only.
    \item {\em Set3}: Limited edge description only.
    \item {\em Set4}: Limited node and edge descriptions.
    \item {\em Set5}: Detailed node and limited edge descriptions.
    \item {\em Set6}: Detailed edge limited node descriptions.
    \item {\em Proposed}: Detailed description for node and edge.
\end{enumerate}
The average number of words for limited descriptions of nodes and edges are 12.5 and 8, respectively, while the average number of words for detailed descriptions of nodes and edges are 42.5 and 20, respectively. 

\begin{figure}[t]
    \centering
    \includegraphics[width=\linewidth]{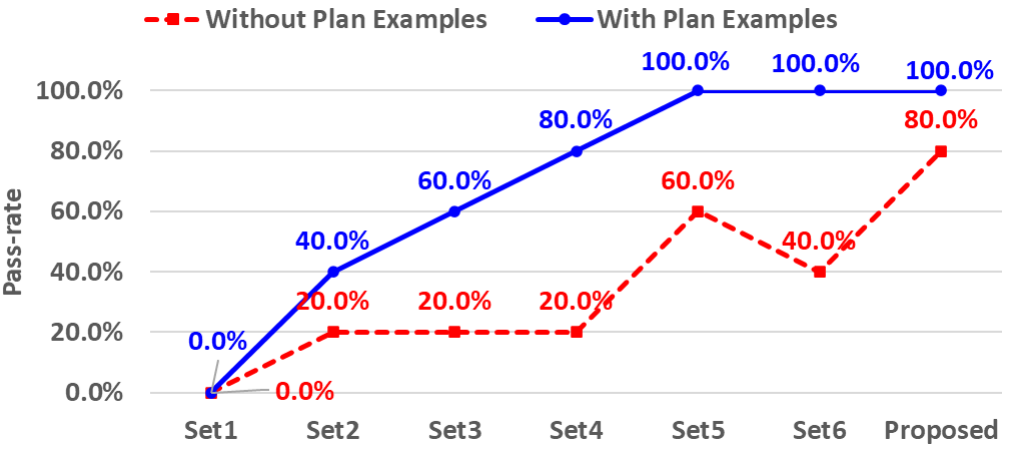} 
    \caption{Pass-rate of varying node and edge descriptions for selected multi-report tasks (i.e., from task M1 to task M5).}
    \label{SensitivityStudyFig}
\end{figure}

Figure~\ref{SensitivityStudyFig} shows the pass-rate of various settings for the node and edge descriptions.
Without plan examples for the multi-report tasks (i.e., the red dashed line), the {\em Proposed} setting achieves an 80\% pass rate. Detailed node descriptions have a significant impact on the pass rate, as seen in the improvements from {\em Set3} to {\em Set5} and from {\em Set6} to {\em Proposed}.
When plan examples are provided (i.e., the blue solid line), the limited edge and node descriptions (i.e., {\em Set4}) also achieve an 80\% pass rate. Adding any detailed information about nodes or edges increases the pass rate to 100\%. 
This sensitivity study highlights the importance of node and edge descriptions in solving multi-report tasks and motivates the continual learning on optimizing the TDRG to further improve multi-report task.


\section{Conclusion} \label{Conclusion}
Our proposed Timing Analysis Agent demonstrates advanced capabilities in hierarchical plan solving and reasoning for complex multi-report tasks from industrial designs through the innovative distilled Timing Debug Relation Graph (TDRG) and coding agentic retrieval methodology. 
Firstly, we show that the proposed flexible and adaptive coding agentic retrieval method not only achieves a 98\% pass rate but also outperforms other retrieval methods by more than 46\% pass-rate on single-report benchmarks.    
Next, we demonstrate that the Timing Analysis Agent achieves a 90\% pass rate on multi-report benchmarks, as evaluated by experienced human engineers. Finally, we have extensively studied the effectiveness and importance of node and edge descriptions within the proposed Timing Debug Relation Graph (TDRG) through a sensitivity analysis of the multi-report benchmarks.







\bibliographystyle{unsrt}
\bibliography{main}

\begin{thebibliography}{10}

\bibitem{yao2022react}
Shunyu Yao, Jeffrey Zhao, Dian Yu, Nan Du, Izhak Shafran, Karthik Narasimhan, and Yuan Cao.
\newblock React: Synergizing reasoning and acting in language models.
\newblock {\em arXiv preprint arXiv:2210.03629}, 2022.

\bibitem{bairi2024codeplan}
Ramakrishna Bairi, Atharv Sonwane, Aditya Kanade, Arun Iyer, Suresh Parthasarathy, Sriram Rajamani, B~Ashok, and Shashank Shet.
\newblock Codeplan: Repository-level coding using llms and planning.
\newblock {\em Proceedings of the ACM on Software Engineering}, 1(FSE):675--698, 2024.

\bibitem{liu2024stall+}
Junwei Liu, Yixuan Chen, Mingwei Liu, Xin Peng, and Yiling Lou.
\newblock Stall+: Boosting llm-based repository-level code completion with static analysis.
\newblock {\em arXiv preprint arXiv:2406.10018}, 2024.

\bibitem{luo2024repoagent}
Qinyu Luo, Yining Ye, Shihao Liang, Zhong Zhang, Yujia Qin, Yaxi Lu, Yesai Wu, Xin Cong, Yankai Lin, Yingli Zhang, et~al.
\newblock Repoagent: An llm-powered open-source framework for repository-level code documentation generation.
\newblock {\em arXiv preprint arXiv:2402.16667}, 2024.

\bibitem{ziletti2024retrieval}
Angelo Ziletti and Leonardo D'Ambrosi.
\newblock Retrieval augmented text-to-sql generation for epidemiological question answering using electronic health records.
\newblock {\em arXiv preprint arXiv:2403.09226}, 2024.

\bibitem{dong2024large}
Haoyu Dong and Zhiruo Wang.
\newblock Large language models for tabular data: Progresses and future directions.
\newblock In {\em Proceedings of the 47th International ACM SIGIR Conference on Research and Development in Information Retrieval}, pages 2997--3000, 2024.

\bibitem{jiang2023structgpt}
Jinhao Jiang, Kun Zhou, Zican Dong, Keming Ye, Wayne~Xin Zhao, and Ji-Rong Wen.
\newblock Structgpt: A general framework for large language model to reason over structured data.
\newblock {\em arXiv preprint arXiv:2305.09645}, 2023.

\bibitem{sui2024table}
Yuan Sui, Mengyu Zhou, Mingjie Zhou, Shi Han, and Dongmei Zhang.
\newblock Table meets llm: Can large language models understand structured table data? a benchmark and empirical study.
\newblock In {\em Proceedings of the 17th ACM International Conference on Web Search and Data Mining}, pages 645--654, 2024.

\bibitem{chang2024data}
Kaiyan Chang, Kun Wang, Nan Yang, Ying Wang, Dantong Jin, Wenlong Zhu, Zhirong Chen, Cangyuan Li, Hao Yan, Yunhao Zhou, et~al.
\newblock Data is all you need: Finetuning llms for chip design via an automated design-data augmentation framework.
\newblock {\em arXiv preprint arXiv:2403.11202}, 2024.

\bibitem{ho2024large}
Chia-Tung Ho and Haoxing Ren.
\newblock Large language model (llm) for standard cell layout design optimization.
\newblock {\em arXiv preprint arXiv:2406.06549}, 2024.

\bibitem{ho2024verilogcoder}
Chia-Tung Ho, Haoxing Ren, and Brucek Khailany.
\newblock Verilogcoder: Autonomous verilog coding agents with graph-based planning and abstract syntax tree (ast)-based waveform tracing tool.
\newblock {\em arXiv preprint arXiv:2408.08927}, 2024.

\bibitem{liu2024layoutcopilot}
Bingyang Liu, Haoyi Zhang, Xiaohan Gao, Zichen Kong, Xiyuan Tang, Yibo Lin, Runsheng Wang, and Ru~Huang.
\newblock Layoutcopilot: An llm-powered multi-agent collaborative framework for interactive analog layout design.
\newblock {\em arXiv preprint arXiv:2406.18873}, 2024.

\bibitem{wu2024eda}
Bing-Yue Wu, Utsav Sharma, Sai Rahul~Dhanvi Kankipati, Ajay Yadav, Bintu~Kappil George, Sai~Ritish Guntupalli, Austin Rovinski, and Vidya~A Chhabria.
\newblock Eda corpus: A large language model dataset for enhanced interaction with openroad.
\newblock {\em arXiv preprint arXiv:2405.06676}, 2024.

\bibitem{wang2024executable}
Xingyao Wang, Yangyi Chen, Lifan Yuan, Yizhe Zhang, Yunzhu Li, Hao Peng, and Heng Ji.
\newblock Executable code actions elicit better llm agents.
\newblock {\em arXiv preprint arXiv:2402.01030}, 2024.

\bibitem{wu2023autogen}
Qingyun Wu, Gagan Bansal, Jieyu Zhang, Yiran Wu, Shaokun Zhang, Erkang Zhu, Beibin Li, Li~Jiang, Xiaoyun Zhang, and Chi Wang.
\newblock Autogen: Enabling next-gen llm applications via multi-agent conversation framework.
\newblock {\em arXiv preprint arXiv:2308.08155}, 2023.

\bibitem{Llama3}
{Meta}.
\newblock {meta-llama/llama3. Original-date: 2024-03-15T17:57:00Z.}
\newblock \url{https://build.nvidia.com/meta/llama3-70b}, 2024.

\bibitem{jones2000probabilistic}
K~Sparck Jones, Steve Walker, and Stephen~E. Robertson.
\newblock A probabilistic model of information retrieval: development and comparative experiments: Part 2.
\newblock {\em Information processing \& management}, 36(6):809--840, 2000.

\bibitem{lewis2020retrieval}
Patrick Lewis, Ethan Perez, Aleksandra Piktus, Fabio Petroni, Vladimir Karpukhin, Naman Goyal, Heinrich K{\"u}ttler, Mike Lewis, Wen-tau Yih, Tim Rockt{\"a}schel, et~al.
\newblock Retrieval-augmented generation for knowledge-intensive nlp tasks.
\newblock {\em Advances in Neural Information Processing Systems}, 33:9459--9474, 2020.

\bibitem{HybridSearch}
{Ashish Abraham, Mór Kapronczay, and Robert Turner}.
\newblock {Optimizing RAG with Hybrid Search \& Reranking}.
\newblock \url{https://superlinked.com/vectorhub/articles/optimizing-rag-with-hybrid-search-reranking}, 2024.

\end{thebibliography}

\end{document}